\documentclass[10pt,letterpaper]{article}
\usepackage{opex3}
\usepackage{epstopdf}
\usepackage{amssymb}
 \usepackage{graphicx}
\usepackage{multirow}
\usepackage{float}
\usepackage{cite}

\begin{document}

\title{Optimized generation of spatial qudits by using a pure phase spatial light modulator}

\author{J.~J.~M. Varga,$^1$ L.~Reb\'on,$^{2,*}$ M.~A.~Sol\'is-Prosser,$^{3,4}$ L.~Neves,$^5$ S.~Ledesma,$^1$ and C.~Iemmi$^1$}

\address{$^1$Departamento de F\'isica, FCEyN, Universidad de Buenos Aires, Buenos Aires 1428, Argentina\\
         $^2$Departamento de F\'isica, IFLP, Universidad Nacional de La Plata, C.C. 67, 1900 La Plata, Argentina\\
$^3$ Center for Optics and Photonics, Universidad de Concepci\'on, Casilla 4016, Concepci\'on, Chile\\
$^4$ Departamento de F\'isica, MSI-Nucleus on Advanced Optics, Universidad de Concepci\'on, Concepci\'on, Chile \\
$^5$ Departamento de F\'isica, Universidade Federal de Minas Gerais, Caixa Postal 702, Belo~Horizonte, MG 30123-970, Brazil}

\email{* rebon@fisica.unlp.edu.ar} 



\begin{abstract}

We present a method for preparing arbitrary pure states of spatial qudits, namely, $D$-dimensional ($D \geq 2$) quantum systems carrying information in the transverse momentum and position of single photons.  For this purpose, a set of $D$ slits with complex transmission are displayed on a spatial light modulator (SLM). In a recent work we have shown a method that requires a single phase-only SLM to control independently the complex coefficients which define the quantum state of dimension $D$. The amplitude information was codified by introducing phase gratings inside each slit and the phase value of the complex transmission was added to the phase gratings. After a spatial filtering process we obtained in the image plane the desired qudit state. Although this method has proven to be a good alternative to compact the previously reported architectures, it presents some features that could be improved. In this paper we present an alternative scheme to codify the required phase values that minimizes the effects of temporal phase fluctuations associated to the SLM where the codification is carried on. In this scheme the amplitudes are set by appropriate phase gratings addressed at the SLM while the relative phases are obtained by a lateral displacement of these phase gratings. We show that this method improves the quality of the prepared state and provides very high fidelities of preparation for any state. An additional advantage of this scheme is that a complete $2\pi$ modulation is obtained by shifting the grating by one period, and hence the encoding is not limited by the phase modulation range achieved by the SLM. Numerical simulations, that take into account the phase fluctuations, show high fidelities for thousands of qubit states covering the whole Bloch sphere surface. Similar analysis are performed for qudits with $D = 3$ and $D =7$.
\end{abstract}




\section{Introduction}

One of the main challenges in the field of quantum information
science is the ability to generate, modify and measure the quantum
systems which are the information carriers in quantum information
processing and computing protocols \cite{BarnettBook}. In this
context, photons are the natural choice for communications since
they are easily transportable, slowly affected by decoherence and
have several degrees of freedom to encode information
\cite{KokBook}. A typical encoding are the so-called spatial qudits,
namely, $D$-dimensional ($D\geq 2$) quantum systems carrying
information in the discretized transverse momentum and position of
single photons \cite{Neves05,Hale05}. In the simplest approach, this
discretization is achieved when the photons are made to pass through
an aperture with $D$ slits which sets the qudit dimension
\cite{Neves04}. Due to this simplicity, spatial qudits enables one
to work in high dimensions without cumbersome optical setups and for
that reason they have drawn interest for miscellaneous applications
such as quantum information protocols \cite{Prosser11}, quantum
games \cite{Kolenderski12}, quantum algorithms \cite{Marques12}, and
quantum key distribution \cite{Etcheverry13}. Most of those
applications has benefited from the recent developments on the
control of spatial qudits based on the technology of electrically
addressed spatial light modulators (SLMs). SLMs have dramatically
simplified and broadened the range of operations that can be
implemented in real time on spatial qudits for state preparation,
transformation and measurement \cite{Lima09, Lima11, Solis13, Solis10}.

Initially, Lima \emph{et al.} \cite{Lima11} have shown that by
imaging the output beam of an amplitude-only SLM onto a phase-only
SLM allows one to get complete and independent control of the
amplitude and phase of the complex coefficients that define the
qudit state. Therefore, this scheme requires two SLMs at each link
of the setup (preparation, transformations, and measurements) in
order to implement arbitrary operations at each one. Besides being
costly in terms of optical resources, it entails two drawbacks: (i)
the overall diffraction efficiency at each link is very low, and
(ii) in order to avoid even more losses, the image of the first SLM
must match the second one pixel by pixel, which is difficult in
practice. In a recent Letter \cite{Solis13}, we presented a
proof-of-principle demonstration of a method which proposed the use
of a single phase-only SLM to control independently the amplitude
and phase of the state coefficients. This method is, thereby, less
costly, offer a much higher diffraction efficiency (we estimated a
10 times higher efficiency), which is relevant when working with
single photon sources, and does not require an optical system for
projecting the image of a SLM onto a second one in order to obtain
the complex modulation. Among the different techniques to represent a complex function in a single SLM \cite{Cohn94, Birch01,vanPutten08} our method follows the proposals of
\cite{Davis99,Bagnoud04} for encoding amplitude and phase
information onto a phase-only SLM. To this end, phase diffraction
gratings were displayed in those zones corresponding to the slits.
The desired amplitude of the complex coefficients, was obtained by
controlling the amount of light diffracted on the first order which
is a function of the phase modulation depth of the grating. The
phase of the complex coefficients, which defines the required
relative phase, was achieved by adding a constant phase value to the
 grating. The required complex light distribution was obtained
after filtering the first diffracted order in
the Fourier plane. We shall refer to this method as the
\emph{phase-addition} (PA) method.

Although the PA method gives, on average, good fidelities of
preparation for spatial qubits and qudits of dimension at least up to 7, we
observed that many of the prepared states had their fidelities
reduced under the same
experimental conditions for state preparation and characterization.
In that paper \cite{Solis13}, we pointed out that this effect was possibly due to
temporal phase fluctuations of the used SLM which was based on liquid crystal on
silicon (LCoS) technology.  In fact, LCoS may lead to flicker in the
optical beam because of the digital addressing scheme (pulse
width modulation) which introduces, among other undesirable effects,
those phase fluctuations \cite{Lizana08,Lizana10} that affect the
quality of the encoded state. This spurious effect will be amplified
as the number of SLMs in a given setup increases, so it is desirable
to eliminate or at least minimize it.

In this paper we first analyze, by numerical simulation, the effects
of the phase fluctuations on the quality of the states prepared by
the PA method and compare the results with the experimental ones
shown in the previous work. Our results corroborate the conjecture
that those fluctuations are the main responsible for the reduction of the fidelities. After that, we
propose an alternative scheme of encoding spatial qudits that minimizes those
effects and, consequently, improve the quality of the preparation.
In this scheme the amplitudes of the slits also are controlled by means of blazed gratings, but the values of the phases of the state coefficients are obtained by performing lateral displacements of the gratings instead of by adding a constant phase.
In this way the required phase is controlled only by the grating
position, which is not affected by phase fluctuations. We shall
refer to this method as the \emph{grating-displacement} (GD) method.
Besides improving the preparation, the new encoding scheme is more
flexible than the previous one regarding the phase modulation range achieved by the SLM. The PA method
requires a SLM with a phase modulation of at least $2\pi$, while the GD method is not limited by this condition
since a complete $2\pi$ modulation is obtained by shifting the grating by one period. This fact is
important specially when long wavelengths (usually near IR), as those obtained by parametric
down-conversion \cite{Burnham70,Kwiat95,Kwiat99,Xue2010}, are used.
The performance of the GD method is analyzed by simulating the
preparation of arbitrary states and their tomographic
reconstruction. We report the results obtained for thousands of
qubit states covering the whole Bloch sphere surface and spatial
qudits of dimension $D=3$ and $D=7$. By studying the fidelities of
preparation we show that the GD method overcomes the PA one.

\section{The grating-displacement method}

The encoding process for the generation of pure states of spatial
qudits can be explained as follows. When a paraxial and
monochromatic single-photon field is transmitted through an aperture
described by a complex transmission function $A(\mathbf{x})$, its
state, assumed here to be pure, is transformed as
\begin{equation}     \label{eq:photon_transmitted}
|\Psi\rangle=\int\!d\mathbf{x}\,\psi(\mathbf{x})|1\mathbf{x}\rangle \;\;\stackrel{A(\mathbf{x})}{\Longrightarrow}
\;\;\int\!d\mathbf{x}\,\psi(\mathbf{x})A(\mathbf{x})|1\mathbf{x}\rangle,
\end{equation}
where $\mathbf{x}=(x,y)$ is the transverse position coordinate and
$\psi(\mathbf{x})$ is the normalized transverse probability
amplitude for this state, i.e.,
$\int\!d\mathbf{x}\,|\psi(\mathbf{x})|^2=1$. Now, let us consider
that $A(\mathbf{x})$ is an array of $D\geq 2$ rectangular slits of
width $2a$, period $d$ and length $L(\gg a,d)$, where each slit,
$\ell$, has a transmission amplitude $\beta_\ell$. Thus,
$A(\mathbf{x})$ will be given by
\begin{equation}    \label{eq:Dslits}
A(\mathbf{x})= {\rm rect}\!\left(\frac{x}{L}\right)\times\sum_{\ell=0}^{D-1}\beta_\ell\;{\rm rect}\!\left(\frac{y-\eta_\ell d}{2a}\right),
\end{equation}
where $\eta_\ell=\ell+(D-1)/2$. Without loss of generality and for
simplicity, we will assume that $\psi(\mathbf{x})$ is constant
across the region of  the slits. Hence, the state of the transmitted
photon in Eq.~(\ref{eq:photon_transmitted}) will be \cite{Neves05}
\begin{equation}     \label{eq:spatial_qudit}
|\psi\rangle=\sum_{\ell=0}^{D-1}\tilde{\beta}_\ell|\ell\rangle,
\end{equation}
where
$\tilde{\beta}_\ell=\beta_\ell/\sqrt{\sum_{j=0}^{D-1}|\beta_j|^2}$
and $|\ell\rangle$ denotes the state of the photon passing through
the slit $\ell$.

As we have shown in \cite{Solis13}, in order to prepare arbitrary
states of the form of Eq.~(\ref{eq:spatial_qudit}) with a single
phase-only SLM, a phase one-dimensional diffraction grating is
displayed on the different regions of the SLM, each of them
corresponding to a particular slit. In this work a blazed phase
profile is selected  to achieve the maximum diffraction efficiency
in the first order, which can be expressed as \cite{GoodmanBook}
\begin{equation}
\epsilon_1={\rm sinc}^2\left(1-\frac{\varphi_0}{2\pi}\right),\label{blazed-grating}
\end{equation}
where $\varphi_0$ is the phase modulation depth and ${\rm
sinc}(u)=\sin(\pi u)/(\pi u)$. When $\varphi_0=2\pi$ the first order
efficiency has a maximum value of 100\%. By selecting other value
for $\varphi_0$, it is possible to modulate the amount of light
diffracted on the order and consequently the amplitude of each
slit. Equation~(\ref{blazed-grating}) corresponds to an ideal blazed
profile with continuous modulation. Nevertheless, given that the
representation of the grating period is carried out through a finite
number of pixels, this imposes a discretization in the phase levels
used to generate the blazed profile. Thus if $N$ is the number of
quantization levels, the maximum efficiency value will be assigned
to the maximum amplitude of the slit coefficients, i.e.,
$|\tilde{\beta}_\ell|=1$ corresponds to $\varphi_0=(N-1)/N\times
2\pi$. Other amplitude values will correspond to other values of
$\varphi_0$ which are obtained from Eq.~(\ref{blazed-grating}). If
the employed SLM do not reach 2$\pi$ phase modulation then the
relative slit amplitudes should be recalculated with respect to the
maximum efficiency achieved. In order to avoid the introduction of
additional phases in the encoding process, the phase gratings should
be designed with zero mean value. However, as the SLM can only
display positive phase values, the gratings are generated with a
mean value equal to half of the maximum phase modulation depth which
is $(N-1)/N\times\pi$ for a blazed grating.

Let us now describe the GD method to control the phase of the
complex coefficients,
${\arg(\tilde{\beta}_\ell)}$, in Eq.~(\ref{eq:spatial_qudit}). It can be understood by analyzing the
transfer function of the grating. If it is assumed that when the
grating is centered at $x=0$ its phase value is zero, the transfer
function of the slit $\ell$ in the far field can be written as:
\begin{equation}
T_{\ell}(x)=\sum_nt_ne^{-i\frac{2\pi}{p}xn},
\end{equation}
where $t_n$ is the amplitude of the $n$-th diffraction order, $x$ is
the position along the grating and $p$ is the grating period, both
measured in pixel units. A lateral displacement of the grating by a
distance of $\delta_{\ell}$ pixels from $x=0$, introduces a phase
shift
 \begin{equation}
\phi_{\ell n}=\frac{2\pi n}{p}\delta_{\ell}
\end{equation}
in the $n$-th order, since the transfer function now is given by
\begin{equation}
T_{\ell}(x-\delta_\ell)=\sum_nt_ne^{-i\frac{2\pi}{p}(x-\delta_{\ell})n}=\sum_nt_ne^{-i\frac{2\pi}{p}x n}e^{i\frac{2\pi}{p}\delta_{\ell} n}.
\end{equation}
By selecting the first diffraction order to obtain the required
complex modulation, the transfer function of the grating will be
\begin{equation}
T^{(1)}_{\ell}(x-\delta_\ell)=t_1e^{-i\frac{2\pi}{p}x}e^{i\frac{2\pi}{p}\delta_{\ell}},
\end{equation}
and the phase shift introduced by the translation is $\phi_{\ell
1}=2\pi\delta_{\ell}/p$. When the maximum amplitude $t_1$ is
normalized, regardless what happens in the other diffraction orders, its value coincides with $\sqrt{\epsilon_1}$ in
Eq.~(\ref{blazed-grating}).

We will show in the next sections that as the phase of the complex
coefficients is determined by the grating position, its value is
almost unaffected by phase fluctuations.

\section{Numerical simulations for state preparation}

\subsection{Proposed experimental setup}

\begin{figure}[H]
\centerline{\includegraphics[width=0.65\columnwidth]{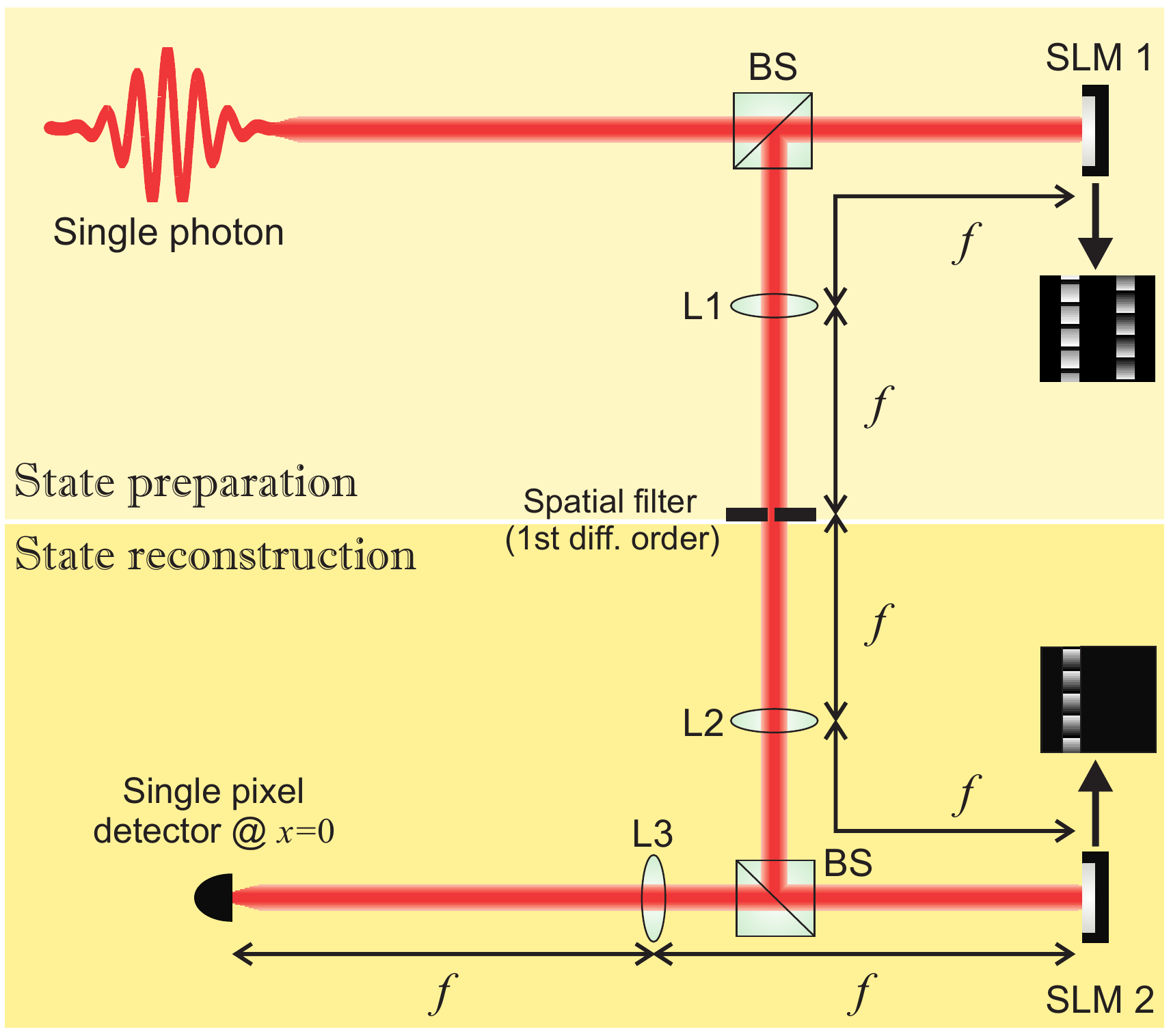}}
\caption{Proposed experimental setup to simulate the preparation and
characterization of spatial qudit states with a single phase-only
SLM subjected to temporal phase fluctuations. BS: beam splitter; L:
lens; SLM: reflective phase-only spatial light modulator; $f$: focal length of the lenses. The insets show examples of phase masks
addressed at the SLMs for state preparation (top) and tomography
(bottom).} \label{fig:setup}
\end{figure}

In order to analyze the performance of both PA and GD encoding
methods against the variation of different parameters of the SLM
(phase fluctuation levels, number of pixels used to represent a
grating period, etc.) we carried out numerical simulations of a
realistic optical setup designed to prepare and characterize spatial
qudit states. The proposed experimental setup is depicted in
Fig.~\ref{fig:setup}. A given source generates a single-photon field 
in the pure state given by the left part of
Eq.~(\ref{eq:photon_transmitted}). As mentioned earlier, we assume
that the transverse probability amplitude $\psi(\mathbf{x})$ is
constant across the region where the slits are displayed in the SLM.
In the upper part of the setup, used for state preparation, a
phase-only SLM (SLM1) is addressed with a phase mask---either by PA
or GD method---corresponding to the spatial qudit state intended to
be prepared. SLM1 is placed in the front focal plane of lens L1 and
an iris diaphragm is placed at its back focal plane in order to
filter the first diffracted order which carries the required
information \cite{Solis13}. In the lower part of the setup, used to
reconstruct the quantum states by tomography, the Fourier transform
of the first diffracted order is projected onto a second SLM (SLM2),
placed at the back focal plane of L2. SLM2 is used to encode the
measurement bases employed to perform the tomographic process as
described in \cite{Lima11}. A single pixel detector is placed at the
Fourier transform plane of SLM2, in the center of its first
diffraction order ($x=0$). This detector registers the single count
rates that after normalization give us the probabilities to
reconstruct the states. It is important to note that with this
configuration it is possible to perform arbitrary projections of the
input state without the need to carry on measurements in the near
field. In this way, the optical setup remains unchanged. In
particular, we perform projections onto the informational complete
set of mutually unbiased basis \cite{Ivanovic81,Wootters89}
following the experiment reported in \cite{Lima11}. Finally, we
apply maximum likelihood technique to obtain the best state
estimation consistent with the requirements of a physical state
\cite{Fiurasek01}.

\subsection{Model for temporal phase fluctuations}
\begin{figure}[H]
\centerline{\includegraphics[width=.85\columnwidth]{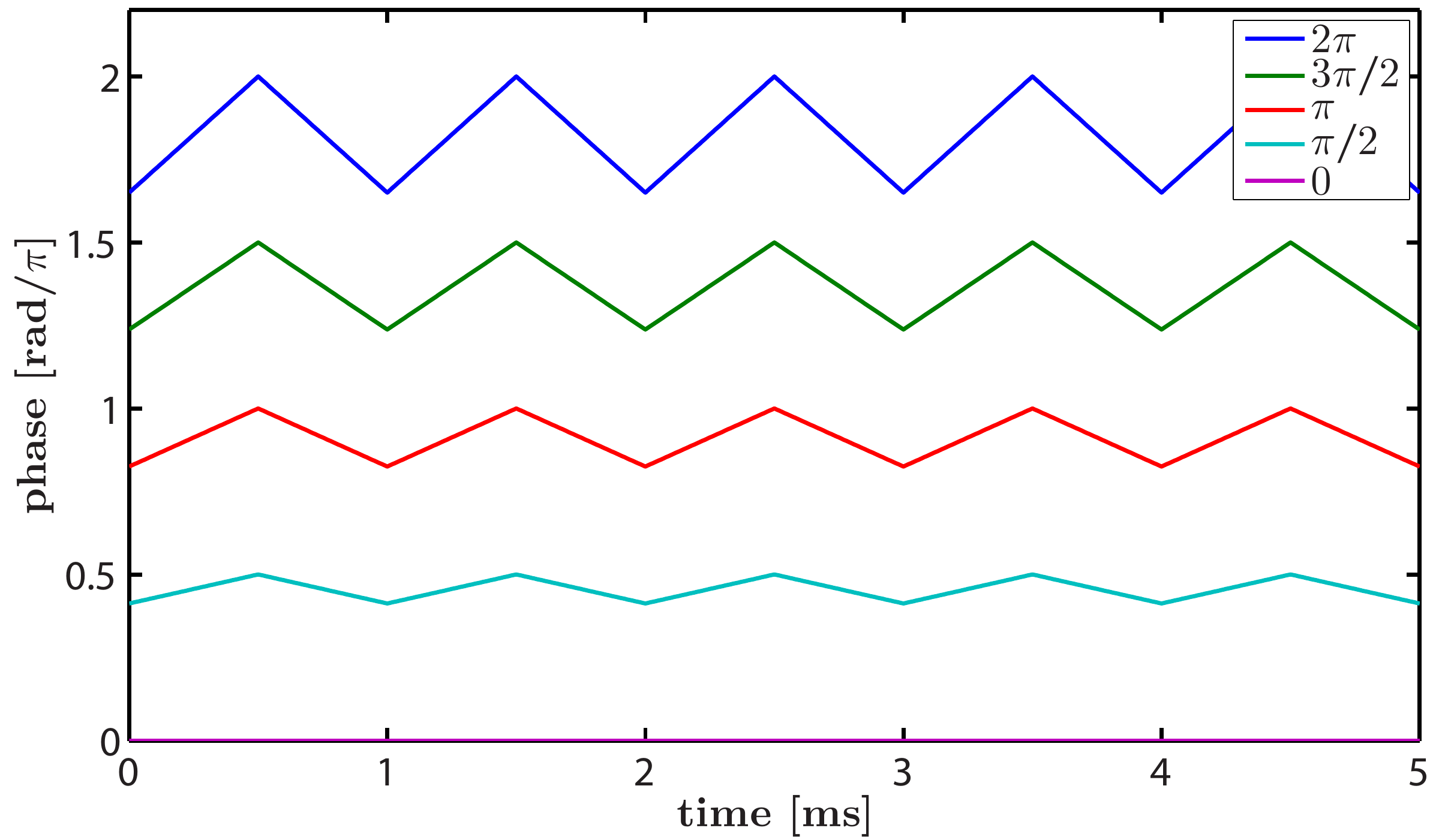}}
\caption{Model for the temporal fluctuations used in the numerical
simulations. Here, the different curves correspond to different phase
modulations addressed on the SLM when the amplitudes fluctuation are $20\%$ of the average phase value.} \label{fig:fluct}
\end{figure}

As it was mentioned, reflective  SLMs such as LCoSs, are usually
employed to obtain pure phase modulation. In a previous paper
\cite{Lizana08, Marquez14}, a simple model has been proposed to describe the
phase fluctuations that are associated with these devices. 
Although
the exact shape of the perturbation varies from one SLM to another one, and  depends on the pulse width
modulation sequence used to address the electrical signal and the
selected gray level, a suitable approximation which describes the general behavior is a triangular phase
fluctuation whose height increases linearly with the phase value.
Also, it should be considered that in many devices the digital
addressing sequence can be programmed. Shorter sequences offer the
possibility of higher repetition rate in a frame period which leads
to a reduction in the flicker amplitude; nevertheless, the number of
addressable phase levels is lower in these cases \cite{Osten07}. According to the
selected sequence the phase fluctuations can reach values as high as
120\% of the average phase value \cite{Lizana10}. Therefore, to
illustrate their effects on each method (PA and GD) we chose
different amplitudes of fluctuation, corresponding to intermediate
sequences, to perform the numerical simulations. We have considered
phase fluctuations ranging from $20\%$ to $60\%$ of the average phase
value. As an example, a model for typical temporal fluctuation is
shown in Fig.~\ref{fig:fluct} where the amplitudes fluctuation are $20\%$ of the average phase value.

It was noted, through the numerical simulations, that the influence of the synchronization between the temporal signals
sent to both modulators (SLMs 1 and 2 in Fig.~\ref{fig:setup}), on
the final result is irrelevant. As a consequence we have decided to consider the
second SLM, employed to perform the tomographic projections, as a
device without fluctuations and transfer the phase fluctuations of
both elements to the first SLM, used to prepare the state.

\begin{figure}[H]
\centerline{\includegraphics[width=0.98\columnwidth]{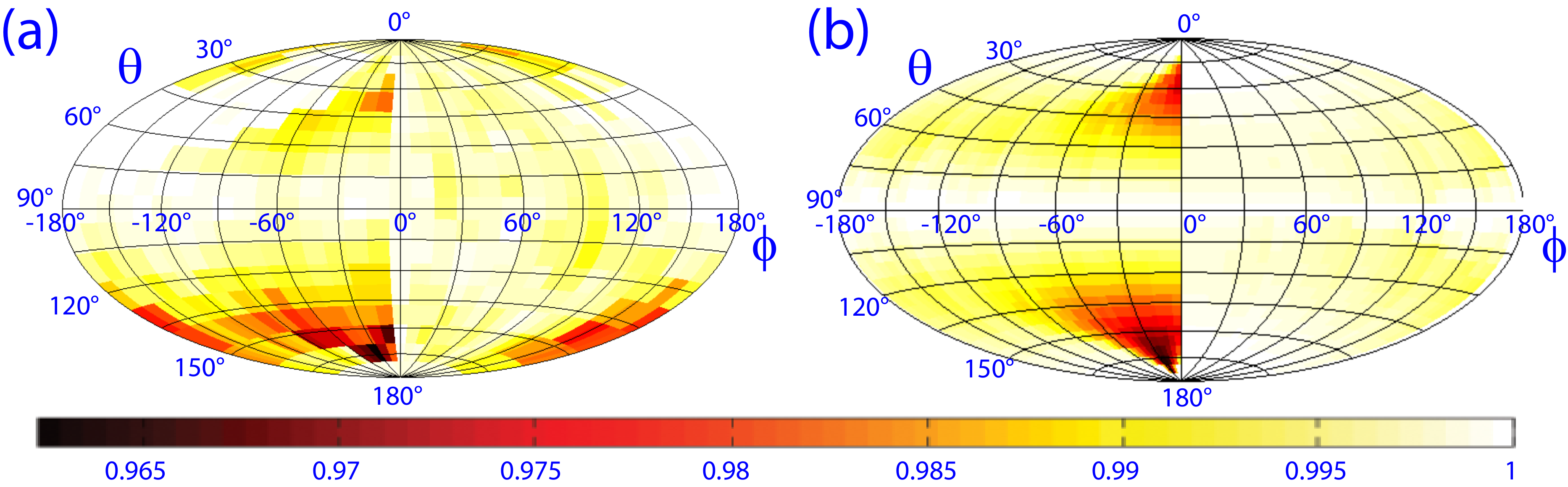}}
\caption{Bloch spheres showing the fidelities of preparation of
spatial qubit states using the phase-addition method: (a)
experimental results from \cite{Solis13} of 561 states uniformly
distributed on the surface; (b) numerical simulations of the
preparation of 2112 states taking into account our model of phase
fluctuations in the SLM (see Fig.~\ref{fig:fluct}). The latitude
$\theta\in[0,\pi]$ and the longitude $\phi\in(-\pi,\pi]$ parametrize
an arbitrary pure state $|\psi\rangle =\cos(\theta/2)|0\rangle +
e^{i\phi}\sin(\theta/2)|1\rangle$ on the sphere surface.}
\label{fig:Bloch-compare}
\end{figure}

In order to validate the mentioned assumptions and the proposed
model, we compared the experimental results obtained in
\cite{Solis13} with the corresponding numerical simulation. To this
end we used the same setup and encoding scheme (PA method) of the
previous work considering the case of a blazed grating. Here and in
the following section, we quantify the quality of the preparation
process with the fidelity,
$F\equiv\langle\psi|\hat{\rho}|\psi\rangle$, between the state
intended to be prepared, $|\psi\rangle$, and the density matrix of
the state actually prepared and reconstructed by tomography,
$\hat{\rho}$. Ideally, it is desirable to have $F=1$. The Bloch
spheres showing the fidelities of preparation of spatial qubits are
shown in Fig.~\ref{fig:Bloch-compare}.
Figure~\ref{fig:Bloch-compare}(a) corresponds to the experimental
results of \cite{Solis13}, and Fig.~\ref{fig:Bloch-compare}(b)
corresponds to the results obtained by numerical simulation.  Although the experimental results show a region 
of low fidelities around $\phi=\pi$ which are not reproduced by the simulation, it is apparent the similarity between them in all other regions of the Bloch's sphere. This slighty difference between both results arises from the fact that the model used to simulate the temporal fluctuations is just a first order aproximation. Besides, the aim of this study is to compare the perfomance of both encoding methods under phase fluctuations and not to exactly reproduce its shape.

\section{Results}

As mentioned above, it is expected that the proposed encoding method
(GD) be barely affected by phase fluctuations, since the phase of
the complex coefficients are determined by the grating positions,
which are not influenced by those fluctuations. Nevertheless, a
possible drawback is that, in principle,  there are fewer available
phase levels than in the PA method. In the latter the available
phase values are determined by the digital sequence used to address
the LCoS whereas in the GD method the phase values are quantized by
the number of pixels used to represent a grating period, {\it i.e.},
the minimum phase shift is $\phi=2\pi/p$. We will show here that for
a sufficiently large value of $p$, this drawback has smaller effects
than the phase fluctuations in the PA method.

\begin{figure}[H]
\centerline{\includegraphics[width=1\columnwidth]{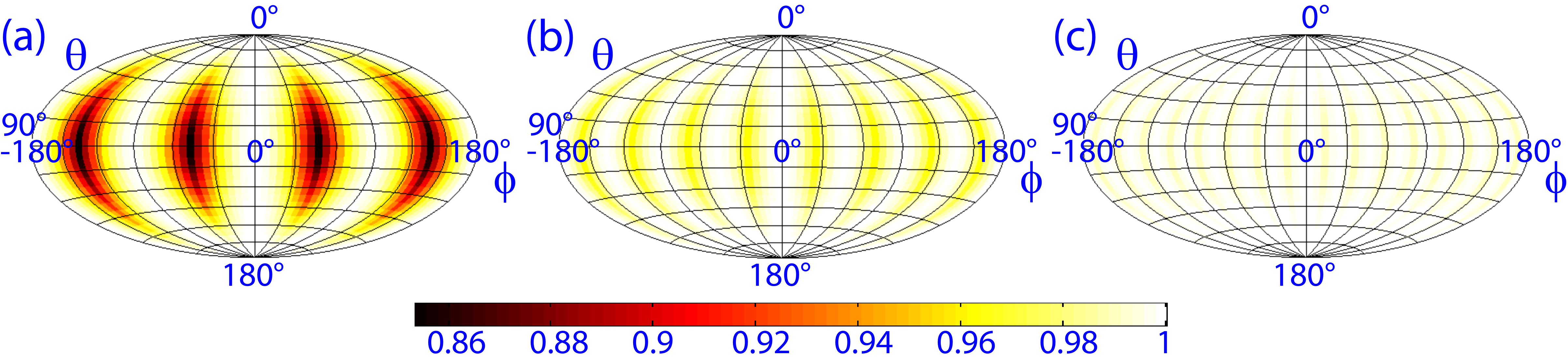}}
\caption{Bloch spheres showing the fidelities of preparation of spatial qubit
states using the grating-displacement method. All graphics were obtained from numerical simulations with 4 (a), 8 (b)
and 16 (c) displacements of the diffraction grating.} \label{fig:Bloch}
\end{figure}

Considering spatial qubits, let us see how the choice of the grating
period, $p$, affects the quality of the preparation. For blazed
gratings with $p=4$, $8$, and $16$ pixels we simulated the preparation
of 2112 states uniformly distributed over the Bloch sphere surface.
The results are shown in
Figs.~\ref{fig:Bloch}(a)-\ref{fig:Bloch}(c), respectively. For each
$p$, it is noticeable a fidelity decrease with a periodic
distribution that depends on how far or close is the phase of the
quantum state from the phase value that we are able to represent.
This decrease is due to the phase quantization rather than the phase
fluctuations introduced by the LCoS and, as expected, it becomes
smaller as $p$ increases. Figure~\ref{fig:Bloch}(a) shows the fidelities obtained with a 4-pixel grating period. In Fig.~\ref{fig:Bloch}(b), with a
$8$-pixel grating period the results are much better and for $p=16$
in Fig.~\ref{fig:Bloch}(c) the obtained fidelities are excellent. In
a realistic scenario, a grating with $16$-pixels period provides
high diffraction efficiency and enables the first diffraction order
to be placed far away from the zeroth order on the Fourier plane. In
this way it can be easily filtered and the light distribution is not
corrupted by unwanted noise. Gratings with larger period will
improve the phase resolution but the first and zeroth order will be
so close that would make it difficult to perform the filtering.
Hence, to compare the performance of both methods (PA and GD)
under fluctuations on the SLMs we have chosen a grating with
16-pixels period.

\vspace{1cm}

\begin{figure}[H]
\centerline{\includegraphics[width=1\columnwidth]{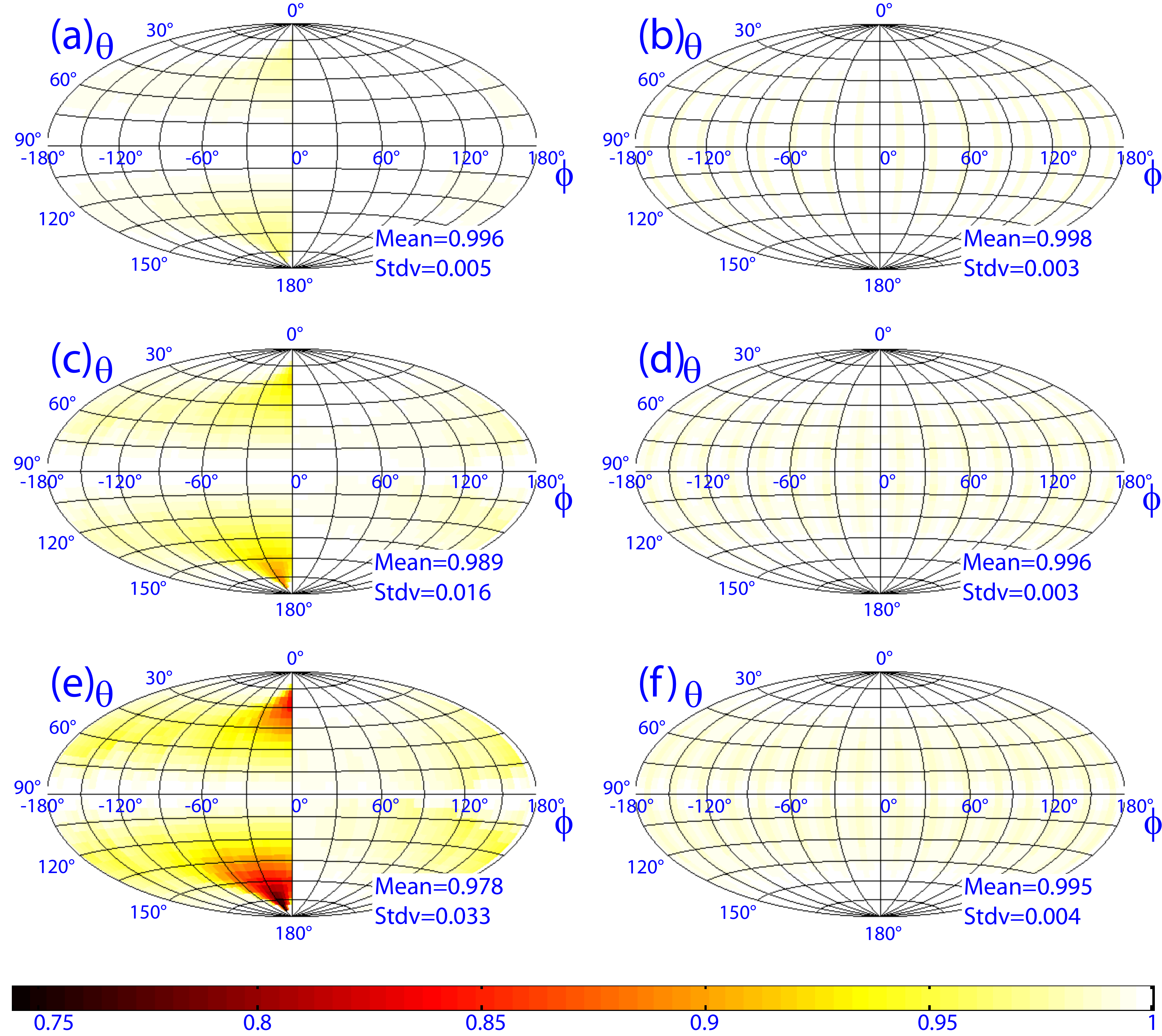}}
\caption{Bloch spheres showing the fidelities of preparation of 2112
spatial qubit states using the PA (first column) and GD (second
column) method. The corresponding phase fluctuation amplitudes are:
20\% (first row), 30\% (second row) and 60\% (third row) of the
average phase value. The insets show the mean fidelity and its
standard deviation.} \label{fig:bloch-lev}
\end{figure}

\begin{figure}[H]

\includegraphics[width=1\linewidth]{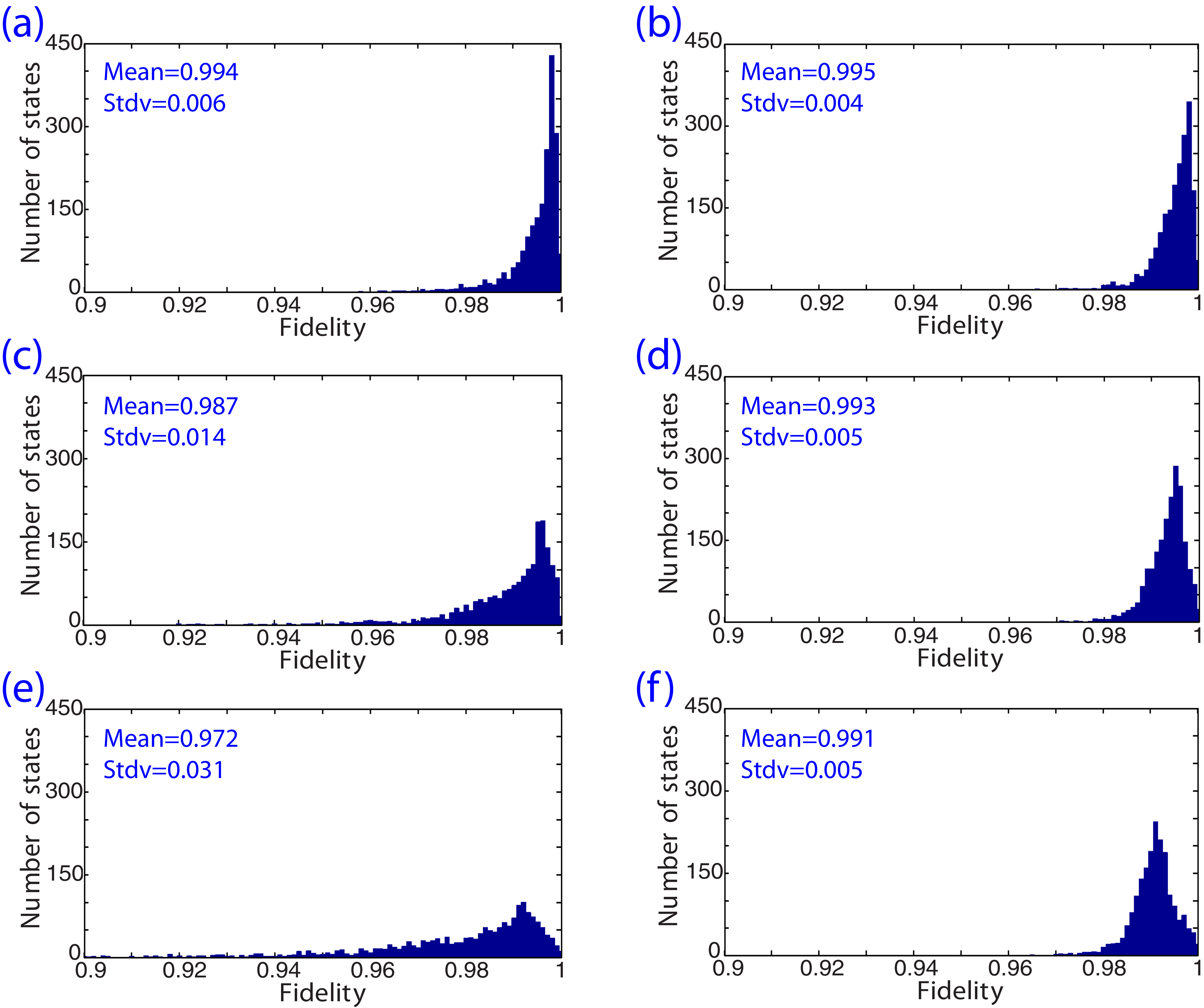}
\caption{Occurrence fidelities of preparation of 2000
spatial qudit states, of dimention $D$=3, using the PA (first column)
and GD (second column) method. The corresponding phase fluctuation
amplitudes are: 20\% (first row), 30\% (second row) and 60\% (third
row) of the average phase value. The insets show the mean fidelity
and its standard deviation.} \label{fig:qudits_3D}
\end{figure}

\begin{figure}[h]

\includegraphics[width=1\linewidth]{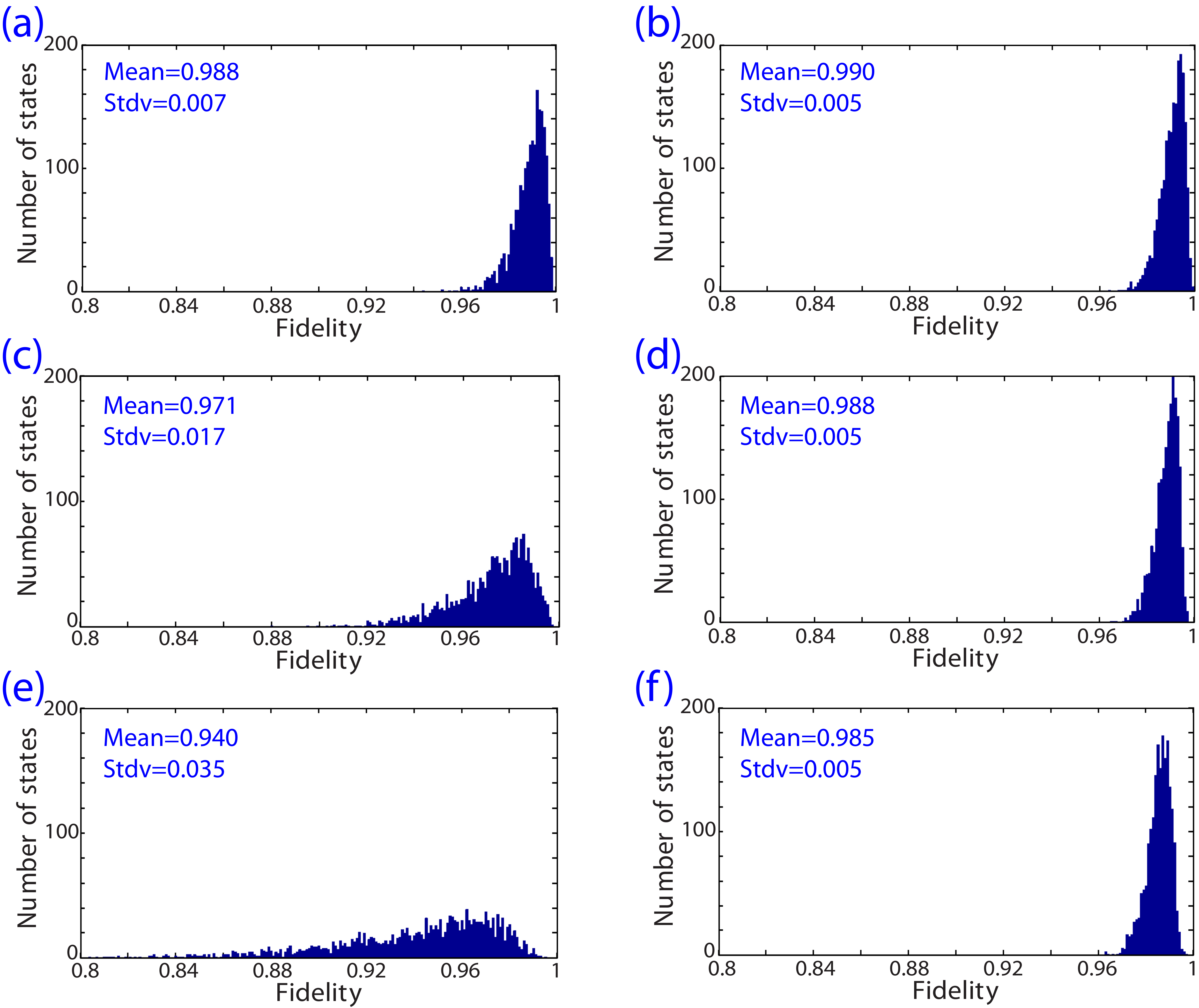}
\caption{Occurrence fidelities of preparation of 2000 spatial qudit states,
of dimension $D$=7, using the PA (first column)
and GD (second column) method. The corresponding phase fluctuation
amplitudes are: 20\% (first row), 30\% (second row) and 60\% (third
row) of the average phase value. The insets show the mean fidelity
and its standard deviation.} \label{fig:qudits_7D}
\end{figure}

Figure~\ref{fig:bloch-lev} shows, for both methods (PA first column
and GD second column), the fidelities of preparation for qubit
states considering three different phase fluctuation amplitudes:
$20\%$  [Figs.~\ref{fig:bloch-lev}(a) and ~\ref{fig:bloch-lev}(b)],
$30\%$ [Figs.~\ref{fig:bloch-lev}(c) and ~\ref{fig:bloch-lev}(d)]
and $60\%$ [Figs.~\ref{fig:bloch-lev}(e) and
~\ref{fig:bloch-lev}(f)] of the average phase value. From these
Bloch spheres it is clear that while the PA method is strongly
dependent on the phase fluctuation amplitude, the GD method remains
almost unaffected. Two thousand arbitrary pure states, corresponding to spatial qudits of higher dimension,
have been prepared. The histograms shown in Fig.~\ref{fig:qudits_3D} represent the number of states of dimension $D=3$ whose fidelity values belong to a particular interval. Figures~\ref{fig:qudits_3D}(a) and~\ref{fig:qudits_3D}(b)
correspond, respectively, to the results obtained  with the PA and GD encoding method for phase fluctuations of 20\% of
the average phase value.  Figures~\ref{fig:qudits_3D}(c) and~\ref{fig:qudits_3D}(d) show the equivalent results for phase fluctuations of 30\%  and  Fig.~\ref{fig:qudits_3D}(e) and~\ref{fig:qudits_3D}(f) the same for fluctuations of 60\%. The results for qudits of dimension $D=7$ are shown in Fig.~\ref{fig:qudits_7D}. Again the left column represents  the histograms obtained from the PA method and the right column to those obtained from the GD method. Figures~\ref{fig:qudits_7D}(a) and \ref{fig:qudits_7D}(b) correspond to  phase fluctuations of 20\%, Figs.~\ref{fig:qudits_7D}(c) and \ref{fig:qudits_7D}(d) to  fluctuations of 30\% and Figs.~\ref{fig:qudits_7D}(e) and \ref{fig:qudits_7D}(f) to  phase fluctuations of 60\%. In a similar way that in the case of a Hilbert space of dimension $D=2$, the results obtained for dimensions $D=3$ and $D=7$ shown that the GD method leads to an increase of the mean fidelity and to a diminution of the standard deviation which agrees with the fact that the encoding applied in the preparation and reconstruction processes is almost not affected by phase noise.

\section{Conclusions}

In this work we have proposed a new method to encode spatial qudit
states using a LCoS as a single phase-only spatial light modulator. In
this method the complex transmissions of the $D$ slits that are used
to represent the quantum state are encoded by means of phase
diffraction gratings. The amplitudes are driven through the phase
modulation depth of the grating and the required phases are
controlled by performing lateral translations of the grating. Given
that the phase values are determined by the grating position the
method is almost unaffected by the phase fluctuations associated to
LCoSs.

We have analyzed the performance of the proposed method by numerical
simulations where we evaluated the preparation of arbitrary states
in a first SLM and its tomographic reconstruction using projective
measurements onto a pre-fixed basis addressed in a second SLM. The proposed method (GD) has been compared with a previous one in \cite{Solis13} where the
required phases were controlled by adding a constant phase value to
the grating (PA). This has been done under different phase
fluctuation intensities for qubits, and qudits of dimension $D=3$
and $D=7$. In all cases the results of the simulations have shown a
greater robustness of the GD method against phase fluctuations 
when comparing with the PA method. In addition the GD method offers wider experimental flexibility allowing one to
use SLMs with a maximum phase modulations below $2\pi$. This feature is
important specially when long wavelengths (usually near IR), as those of photons obtained by parametric down-conversion, are used.

Therefore, the method proposed here may become a valuable tool for experiments based on spatial qudits, specially when one has to assemble two or more SLMs (e.g., for state preparation and transformations) subjected to phase fluctuations. By using the GD method one minimizes the noise effects caused by the fluctuations and, consequently, improve the realization of the protocol of interest.

\section*{Acknowledgments}

This work was supported by UBACyT 20020100100689, CONICET PIP
112-200801-03047 and ANPCYT PICT 2010-02179 (Argentina), CNPq 485401/2013-4 and FAPEMIG APQ-00149-13 (Brazil), and CONICYT PFB08-24 and Milenio ICM P10-030-F (Chile). M.A.S.P. acknowledges financial support from CONICYT (Chile).

\end{document}